\documentclass[letterpaper,english,preprint, prd]{revtex4-1}
\usepackage[T1]{fontenc}
\usepackage[latin9]{inputenc}
\usepackage{babel}
\usepackage{calc}
\usepackage{amsmath}
\usepackage{amssymb}
\usepackage{graphicx}
\usepackage{tikz}
\usepackage{array}
\usepackage{float}
\usepackage{multirow}
\usepackage{eqnarray,amsmath}
\usetikzlibrary{patterns}
\usetikzlibrary{plotmarks}
\usepackage[unicode=true]
 {hyperref}
\usepackage{breakurl}	
\usepackage[compat=1.1.0]{tikz-feynman}
	\tikzset{
	photon/.style={decorate, decoration={snake}, draw=red},
	particle/.style={draw=blue, postaction={decorate},
		decoration={markings,mark=at position .5 with {\arrow[draw=blue]{>}}}},
	antiparticle/.style={draw=blue, postaction={decorate},
		decoration={markings,mark=at position .5 with {\arrow[draw=blue]{<}}}},
	gluon/.style={decorate, draw=black,
		decoration={coil,amplitude=4pt, segment length=5pt}}
}
\makeatletter

\providecommand{\LyX}{\texorpdfstring%
  {L\kern-.1667em\lower.25em\hbox{Y}\kern-.125emX\@}
  {LyX}}
\makeatother

\begin{document}

%\preprint{APS/123-QED}

\title{Excited Quarks Production at FCC and S$pp$C $pp$ Colliders }

\author{Mehmet Sahin}
\email{mehmet.sahin@usak.edu.tr}

\affiliation{Department of Physics, Usak University, Usak, Turkey}

\author{Güral Aydin}
\email{gaydin@mku.edu.tr}

\affiliation{Department of Physics, Hatay Mustafa Kemal University, Hatay, Turkey }

\author{Yusuf Oguzhan Günaydin}
\email{yusufgunaydin@gmail.com}

\affiliation{Department of Physics, Kahramanmaras Sütcü Imam University, Kahramanmaras,
Turkey }

\date{\today}
\begin{abstract}
Potential discovery, observation and exclusion limits of excited $u$ and $d$ quarks with gamma+jet final state are researched at the multi-TeV scale colliders, FCC and S$pp$C in this work.    Both colliders, FCC and S$pp$C, show that excited $u$ and $d$ quarks could be discovered up to 42.1 TeV and 55.2 TeV for $u^\star$, 30.3 TeV and 39.4 TeV for $d^\star$ and 42.3 TeV and 55.5 TeV mass values for degenerate case (m$_{u^\star}$ = m$_{d^\star}$), respectively. Determination of excited quarks compositeness scale  is examined, which will be up to multi-PeV level. Beside these analysis, free parameters ($f_s,\;f\;\text{and}\; f^\prime$) are scanned from around 0.1 up to 1 that show excited quark could be discovered at dozens of TeV with even coupling constants under 0.1.   
%\begin{description}
%\item [{Usage}] Secondary publications and information retrieval purposes.{\small \par}
%\item [{PACS~numbers}] May be entered using the environment \textsf{PACS~numbers}.{\small \par}
%\item [{Structure}] You may use the \texttt{Description} environment to
%structure your abstract.{\small \par}
%\end{description}
\end{abstract}

\pacs{33.15.Ta}

\keywords{Suggested keywords}

\maketitle

\section{\label{sec:intro}Introduction}

The historical development of the elementary particles \cite{sahin2011}
suggest that there could be more fundamental particles than Standard
Model (SM) fermions. Therefore, over the last four decades, both experimental and theoretical particle
physicists have proposed new models 
 claiming that SM fermions could have composite
substructure which are called preons \cite{pati1974,shupe1979,harari1979,terazawa1980,fritzsch1981,terazawa1982,terazawa1983,eichten1983,dsouza1992,celikel1998,desouza2008,fritzsch2016}.
Preonic interactions bring into such new hypothetical particles like
diquarks, dileptons, leptogluons, leptoquarks, excited quarks ($Q^{\star}$)
and excited leptons ($l^{\star}$). As the matter of fact in the SM, there should be
three possible families for excited fermions. This paper focuses
on analysis of the first family spin-1/2 excited quarks ($u^{\star}$
and $d^{\star}$). Existence of the excited quark will be strong verification
of such a state that composite structure of the fundamental particle
might be possible. Excited quark mass is expected much heavier
than the SM quark mass. 

The LHC working plan will be ended in the 2020s and
the more energetic particle colliders are required to observe and
discover this unstable and heavy particles. After that, two new energy
frontier particle collider machines, Future Circular Collider (FCC)
at CERN and Super Proton Proton Collider (S$pp$C) at China, are planned
for the deep investigation on hypothetical particles. FCC working plan will be carried out in two phases. It is expected to reach its final integrated luminosity as 2.5 ab$^{-1}$ at Phase I in 10 years and 15 ab$^{-1}$ at Phase II in 15 years. Overall, integrated luminosity of the FCC is expected as 17.5 ab$^{-1}$ in 25 years \cite{Benedikt:2018csr}. On the other hand, S$pp$C is presumed to reach 22.5 ab$^{-1}$ in 15 years \cite{su2016}. 

There are some experimental \cite{cdf1995,d0_2006,abdallah2006,cms2011,cms2014,cms2016,cms2016photon,atlas2016photon,atlas2016,cms2017,cms2017heavy,atlas2017,sirunyan2018,sirunyan2018gammajet,sirunyan2018WjetZjet}
and phenomenological \cite{low1965,pati1974,terazawa1977,harari1979,shupe1979,terazawa1980,fritzsch1981,terazawa1982,eichten1983,lyons1983,renard1983,terazawa1983,kuhn1984,pancheri1984,rujula1984,hagiwara1985,kuhn1985,baur1987,spira1989,baur1990,jikia1990,dsouza1992,boudjema1993,celikel1998,cakir1999,cakir2000,cakir2001,eboli2002,cakir2004,ccakir2004,cakir2008,harris2011,biondini2012,terazawa2014,terazawa2015,biondini2015,fritzsch2016,panella2017, caliskan2017,caliskan2017a,desouza2008,gunaydin2018,Kaya2018,caliskan2018,Akay_2019}
research that investigate excited fermions and provide mass limits of excited
quarks based on final state particles. Currently, CERN LHC collaborations
(ATLAS and CMS) hold experimental exclusion limits on excited quark
mass in proton-proton collisions as m$_{q^{\star}}=6.0$ TeV at $jj$,
m$_{q^{\star}}=5.5$ TeV at $\gamma j$ final states according to
both experiments, and m$_{q^{\star}}=3.2$ TeV at $Wj$ and m$_{q^{\star}}=2.9$
TeV at $Zj$ final states for the ATLAS experiment, m$_{q^{\star}}=5.0$
TeV at $Wj$ and m$_{q^{\star}}=4.7$ TeV at $Zj$ final states for
the CMS experiment. 

In this research, production cross section of spin-1/2 excited $u$ and $d$ quark ($u^{\star}$ and $d^{\star}$) decaying into gamma+jet final states are analyzed at the FCC and the S$pp$C with their expected integrated luminosities. To do so, interaction Lagrangian of spin-1/2 excited quark and its decay widths are presented in following section \ref{sec:intLag}.  Then,  cross section values in section \ref{sec:cs}, signal-background analysis to determine cuts in section \ref{sec:dist},  and attainable mass, compositeness scale ($\Lambda$) limits, free parameters' scan and conclusions in section \ref{sec:conc} are introduced. 

\section{\label{sec:intLag}Interaction Lagrangian of Spin 1/2 Excited Quarks and Decay Widths}

The first generation SM and excited quarks' isospin structure are described by \cite{baur1990}

\[ \begin{bmatrix} 
u \\
d 
\end{bmatrix}_L , 
\begin{array}{c}
u_R \; d_R, 
\end{array}
\begin{bmatrix} 
u^{\star} \\
d^{\star} 
\end{bmatrix}_L , 
\begin{bmatrix} 
u^{\star} \\
d^{\star} 
\end{bmatrix}_R. \]
Interaction Lagrangian of spin-1/2 excited quarks, which interact with SM gauge bosons and quarks, is presented in Eq. \ref{eq:lagrangian}.  Here, $\Lambda$ represents compositeness scale, $Q_{R}^{\star}$ is right-handed excited quark, $Q_L$ is SM quark, $g_{s},\; g\;\text{and}\;g^{'} $ are gauge coupling constants, $F^a_{\mu \nu}$, $\overrightarrow{W}_{\mu \nu}$, $B_{\mu \nu}$ are the field strength tensors for gluon, SU(2) and U(1), $\lambda^a$ are 3$\times$3 Gell-Mann  matrices,  $\vec{\tau}$ is the Pauli spin matrices, $Y =1/3$  is weak hypercharge and  $f_s$, $f$ and $f'$ are  free parameters. Eq. \ref{eq:lagrangian} have been implemented to CalcHEP \cite{calchep2013}  simulation software via LanHEP \cite{LanHEP,lanhep2016}.

\begin{equation}
L_{eff}=\frac{1}{2\Lambda}\overline{Q_{R}^{\star}}\;\sigma^{\mu\nu}[g_{s}f_{s}\frac{\lambda_{a}}{2}G_{\mu\nu}^{a}+gf\frac{\overrightarrow{\tau}}{2}\overrightarrow{W}_{\mu\nu}+g^{'}f^{'}\frac{Y}{2}B_{\mu\nu}]Q_{L}+h.c.\label{eq:lagrangian}
\end{equation}

The analytical formulae for excited up- and down-quark partial decay width are presented in Eq. \ref{eq:decay}, \ref{eq:decay2}, \ref{eq:decay3}.  In these equations, SM quark mass is neglected because the experimental limits on excited quark mass are at TeV scales \cite{atlas2016photon,cms2016photon}:   
\begin{eqnarray}
\label{eq:decay}
\Gamma_{(Q^{\star}\rightarrow Qg)}  &=&  \frac{1}{3}\alpha_{s}f_{s}^{2}\frac{m_{Q^{\star}}^{3}}{\Lambda^{2}}\\
\label{eq:decay2}
\Gamma_{(Q^{\star}\rightarrow Q\gamma)} &=& \frac{1}{4}\alpha f_{\gamma}^{2}\frac{m_{Q^{\star}}^{3}}{\Lambda^{2}} \\
\label{eq:decay3}
\Gamma_{(Q^{\star}\rightarrow QV)}   &=&  \frac{1}{32\pi}g_{V}^{2}f_{V}^{2}\frac{m_{Q^{\star}}^{3}}{\Lambda^{2}}\left(1-\frac{m_{V}^{2}}{m_{Q^{\star}}^{2}}\right)^{2}\left(2+\frac{m_{V}^{2}}{m_{Q^{\star}}^{2}}\right)
\end{eqnarray}

where, $V$ denotes  $W\;\text{or}\; Z$ bosons, $Q$ represents $u\; \text{or}\; d$ quarks and  $Q^{\star}$ symbolizes  $u^{\star}\; \text{or}\; d^{\star}$ excited quarks and then considering $T_3$ is a third component of weak isospin and $Y$ is a weak hyper charge of an excited quark; $f_{\gamma}=f T_{3}+f^{\prime}(Y / 2)$, $f_{Z}=f T_{3} \cos ^{2} \theta_{W}-f^{\prime}(Y / 2) \sin ^{2} \theta_{W}$, $f_{W}=f / \sqrt{2}$.  

\begin{figure}[t]
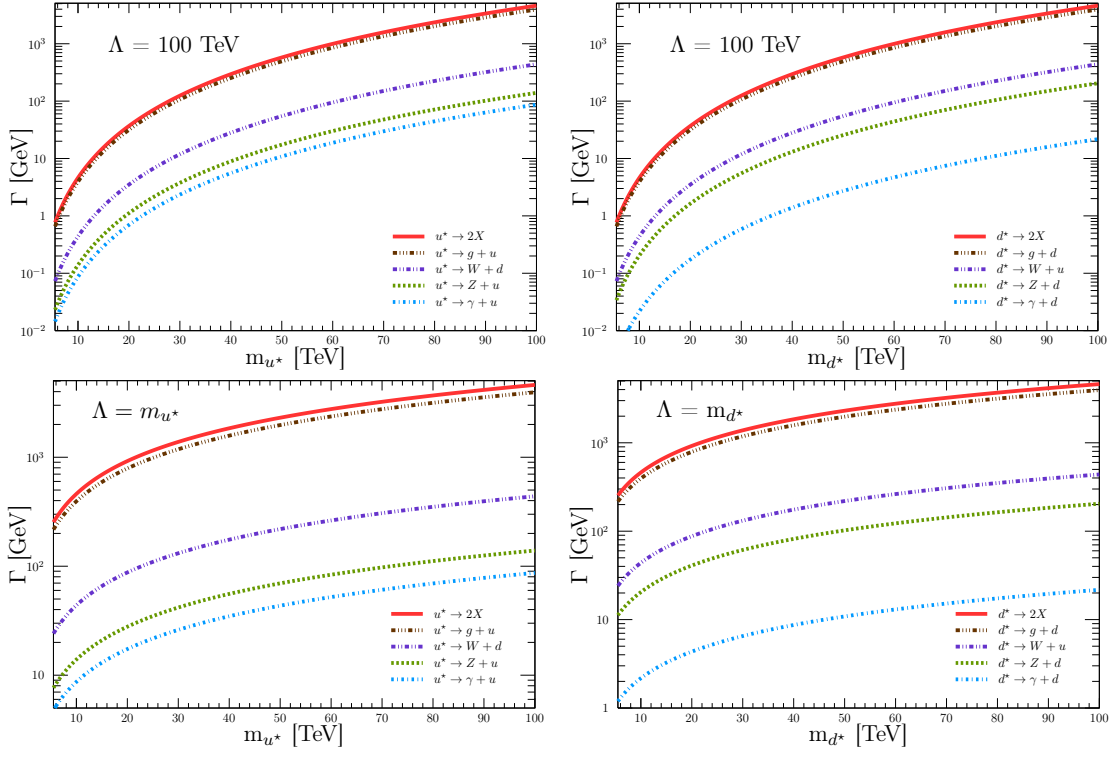

	\scalebox{0.4}{\input{dw_ustar.tex}}	
	\scalebox{0.4}{\input{dw_dstar.tex}}	
	\scalebox{0.4}{\input{dw_ustar_LEM.tex}}	
	\scalebox{0.4}{\input{dw_dstar_LEM.tex}}	
	\caption{Partial and total decay widths for $u^\star$ (left panel) and $d^\star$ (right panel).\label{fig:decay}}
\end{figure}

The decay width values obtained via CalcHEP are summarized in Fig \ref{fig:decay}. Total and partial decay widths for $u^\star\; \text{and}\; d^\star$ are examined starting from 5.5 TeV to 100 TeV mass ranges with $f=f_{s}=f^{\prime}=1$. Besides two different flavors used in analysis, decay widths are plotted in terms of different compositeness values. In Fig. \ref{fig:decay}, compositeness scale was chosen 100 TeV at the first row and it was selected equal to excited quark mass at the second row. 
It is apparent from Fig. \ref{fig:decay} that  $Q^\star\rightarrow \gamma +Q$ channel makes the lowest contribution to total decay width unlike $Q^\star\rightarrow g +Q$ channel. While the latter channel corresponds to dijet final states, the former channel corresponds to gamma+jet final state which is studied in this work.  

\section{\label{sec:cs}Excited Quarks Production Cross Sections at Multi-TeV Scale}

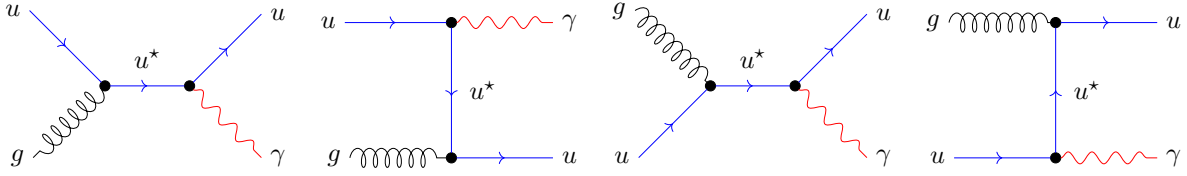
\begin{figure}[t]
\scalebox{0.9}{	\begin{tikzpicture}
		\begin{feynman}
		\coordinate[label=left:$u$] (e1);
		\coordinate[below right=of e1,dot] (aux1);
		\coordinate[below left=of aux1,label=left:$g$] (e2);
		\coordinate[right=1.25cm of aux1,dot] (aux2) {};
		\coordinate[above right=of aux2,label=right:$u$] (e3);
		\coordinate[below right=of aux2,label=right:$\gamma$] (e4);
		
		\draw[particle] (e1) -- (aux1);
		\draw[gluon] (aux1) -- (e2);
		\draw[particle] (aux2) -- (e3);
		\draw[photon] (aux2) -- (e4);
		\draw[particle] (aux1) -- node[label=above:$u^{\star}$] {} (aux2);
		\end{feynman}
		\end{tikzpicture}}
\scalebox{0.9}{\begin{tikzpicture}
\begin{feynman}
		\coordinate[label=left:$u$] (e1);
		\coordinate[right=of e1,dot] (aux1);
		\coordinate[right =of aux1,label=right:$\gamma$] (e2);
		\coordinate[below=2.0cm of aux1,dot] (aux2);
		\coordinate[right=of aux2,label=right:$u$] (e3);
		\coordinate[left=of aux2,label=left:$g$] (e4);
		
		\draw[particle] (e1) -- (aux1);
		\draw[photon] (aux1) -- (e2);
		\draw[particle] (aux2) -- (e3);
		\draw[gluon] (e4) -- (aux2);
		\draw[antiparticle] (aux2) -- node[label=right:$u^{\star}$] {} (aux1);
		\end{feynman}
		\end{tikzpicture}}
\scalebox{0.9}{	\begin{tikzpicture}
	\begin{feynman}
	\coordinate[label=left:$g$] (e1);
	\coordinate[below right=of e1,dot] (aux1);
	\coordinate[below left=of aux1,label=left:$u$] (e2);
	\coordinate[right=1.25cm of aux1,dot] (aux2) {};
	\coordinate[above right=of aux2,label=right:$u$] (e3);
	\coordinate[below right=of aux2,label=right:$\gamma$] (e4);
	
	\draw[gluon] (e1) -- (aux1);
	\draw[particle] (e2) -- (aux1);
	\draw[particle] (aux2) -- (e3);
	\draw[photon] (aux2) -- (e4);
	\draw[particle] (aux1) -- node[label=above:$u^{\star}$] {} (aux2);
	\end{feynman}
	\end{tikzpicture}}
\scalebox{0.9}{	\begin{tikzpicture}
	\begin{feynman}
	\coordinate[label=left:$g$] (e1);
	\coordinate[right=of e1,dot] (aux1);
	\coordinate[right =of aux1,label=right:$u$] (e2);
	\coordinate[below=2.0cm of aux1,dot] (aux2);
	\coordinate[right=of aux2,label=right:$\gamma$] (e3);
	\coordinate[left=of aux2,label=left:$u$] (e4);
	
	\draw[gluon] (e1) -- (aux1);
	\draw[particle] (aux1) -- (e2);
	\draw[photon] (aux2) -- (e3);
	\draw[particle] (e4) -- (aux2);
	\draw[particle] (aux2) -- node[label=right:$u^{\star}$] {} (aux1);
	\end{feynman}
	\end{tikzpicture}}
\caption{Illustration of  signal process Feynman diagrams for direct and indirect production at $pp$ colliders.\label{fig:feynman}}
\end{figure}

\begin{figure*}[t]
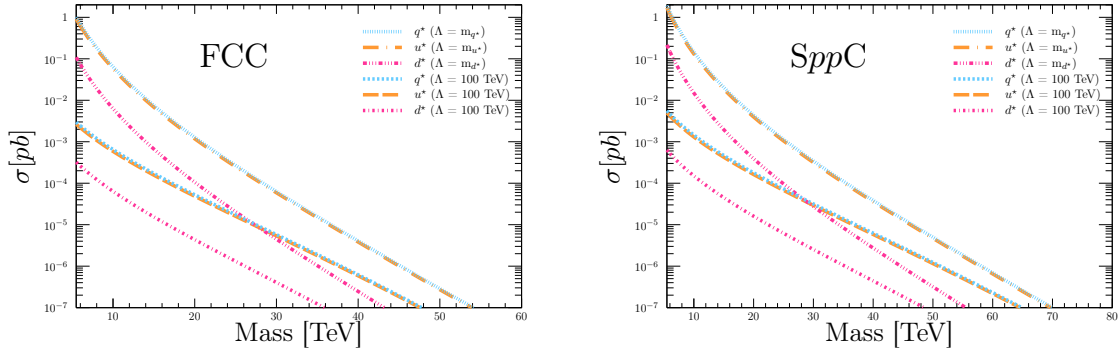

	\scalebox{0.37}{\input{pp_Aj_FCC_Cross_Sections.tex}}
	\scalebox{0.37}{\input{pp_Aj_SppC_Cross_Sections.tex}}	
	\caption{$q^\star$, $u^\star$  and $d^\star$  cross section values for the FCC (left panel) and S$pp$C (rightpanel). \label{fig:cs}}	
\end{figure*}

In this paper, three different cases as the signal process, which are listed in Eq. \ref{eq:signals} are considered to be studied where $q^{\star}$  means $u^{\star} + d^{\star}$. Firstly, production cross section values are computed via CalcHEP simulation software in which cross section plots (Fig. \ref{fig:cs}) are generated by using CTEQ6L1 \cite{pumplin2002,stump2003} quark distribution function and taking renormalization and factorization scales equal to excited quark masses. At proton-proton colliders, resonance production of excited quarks cross section values obtained at LO level are as high as next-to-leading order (NLO) level cross section values due to QCD interactions. Therefore, LO level cross section values are used in experimental studies without computation of any $k$-factor \cite{cmsNOTE2006, cms2011, harris2011} for excited quarks. So, all calculations are done at a leading order (LO) level. According to LHC experiments \cite{atlas2016photon, atlas2017} using different selections of PDF, factorization and renormalization scales  cause less than 1\% contribution on systematic uncertainties for $Q^{\star}\rightarrow \gamma Q$ channel. Figure \ref{fig:feynman} illustrates four Feynman diagrams that contribute to signal cross section for the case described in Eq. \ref{eq:signalsa}. Besides these diagrams, cases described in Eq. \ref{eq:signalsb} and \ref{eq:signalsc} contribute to signal cross section with four and eight Feynman diagrams, respectively. 

\begin{subequations}
\label{eq:signals}
\begin{align}
\label{eq:signalsa}
\text{m}_{u^\star} &> \text{m}_{d^\star} \;\text{for}\; pp\rightarrow u^{\star}+X\rightarrow \gamma u+X\\
\label{eq:signalsb}
\text{m}_{d^\star} &> \text{m}_{u^\star} \;\text{for}\; pp\rightarrow d^{\star}+X\rightarrow \gamma d+X\\
\label{eq:signalsc}
\text{m}_{u^\star} &= \text{m}_{d^\star}\; \text{for}\; pp\rightarrow q^{\star}+X\rightarrow \gamma q+X \;\text{(degenerate state)} 
\end{align}
\end{subequations}

As shown in Fig. \ref{fig:cs}, excited $u$-quark cross section values  are significantly  higher than excited $d$-quark due to proton's two valance up quarks and electric charge difference of excited up (2/3) and down (-1/3) quarks. It is roughly estimated that for 10 events, $u^\star$ and $d^\star$ can be produced up to 46 TeV, 36 TeV mass limits at FCC and 58 TeV and 45 TeV at S$pp$C, respectively. 

\section{\label{sec:dist}Signal Background Analysis of Excited Quarks}

 As it is declared in previous section, three signal processes, $pp\rightarrow u^{\star}+X\rightarrow \gamma u+X$, $ pp\rightarrow d^{\star}+X\rightarrow \gamma d+X$ and $pp\rightarrow q^{\star}+X\rightarrow \gamma q+X \;\text{(degenerate state)}$ are analyzed along with background process $pp \rightarrow \gamma j + X$.  Here, $j$ indicates $u, \, \bar{u},\, d,\,\bar{d},\, c,\, \bar{c},\, s, \,\bar{s},\, b,\, \bar{b}$ and $g$. Considering four different signal mass values with gamma+jet final state, transverse momentum and pseudo-rapidity distributions that depending on normalized differential cross sections are plotted for both colliders. 
 
  \begin{figure*}[h!]
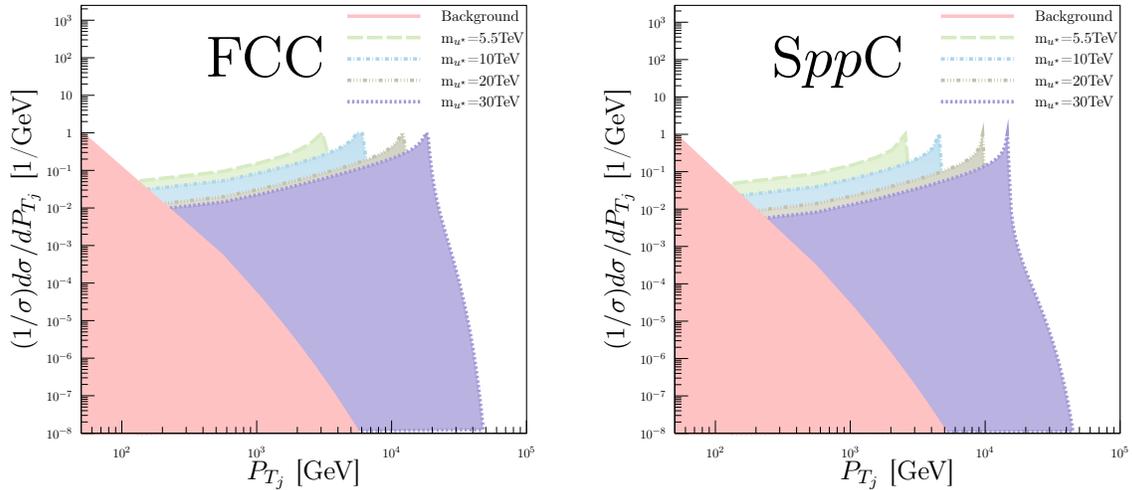

 	\scalebox{0.37}{\input{dist_FCC_pTplot_Aj.tex}}	
 	\scalebox{0.37}{\input{dist_SppC_pTplot_Aj.tex}}		
 	\caption{$u^\star$  transverse momentum distribution plots for the FCC (left panel) and SppC (right panel).  \label{fig:pT_dist}}
 \end{figure*}

\begin{figure*}[b]
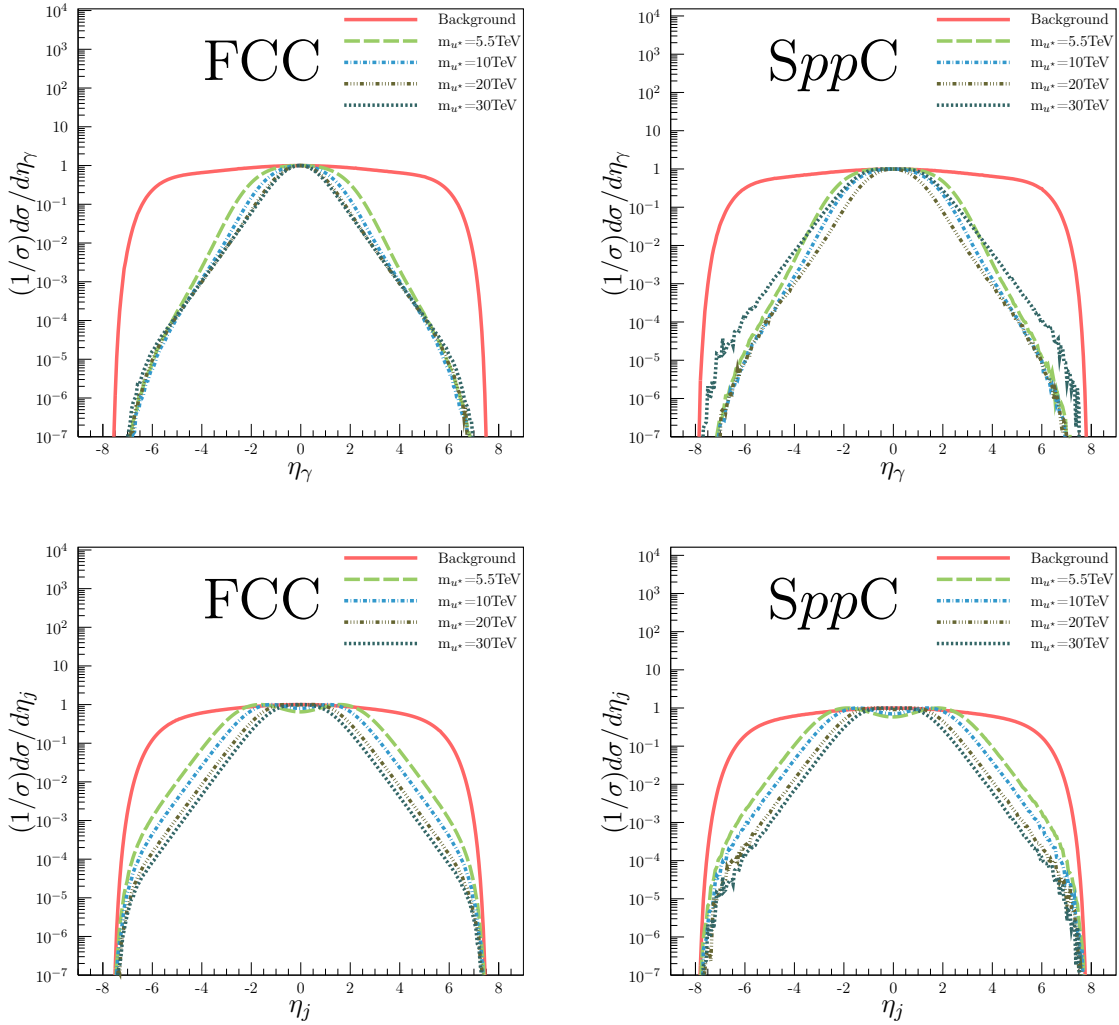

	\scalebox{0.37}{\input{dist_FCC_eTaplot_A.tex}}	
	\scalebox{0.37}{\input{dist_SppC_eTaplot_A.tex}}	
	\scalebox{0.37}{\input{dist_FCC_eTaplot_j.tex}}	
	\scalebox{0.37}{\input{dist_SppC_eTaplot_j.tex}}	
	\caption{$u^\star$  pseudo-rapidity distribution plots for the FCC (left panel) and S$pp$C (right panel)  at $\gamma$ (first row) and $jet$ (second row) final states. \label{fig:eta_dist}}
\end{figure*}

 Transverse momentum ($P_T$), pseudo-rapidity ($\eta_j\;\text{or}\; \eta_{\gamma}$) and invariant mass (m$_{\gamma j}$) cuts that distinguish the signal process from the background process are implemented. Both final state particle's transverse momentum distributions show almost the same characteristic. For this reason, one distribution plot is presented for illustration. As shown in Fig. \ref{fig:pT_dist}, 2000 GeV cut significantly reduces the background while the signal remains almost unchanged. This cut is far beyond the experimental $P_T$ cut because if $P_T > 150$ GeV and events contain at least one photon and one jet candidate,  photon and jet registration is nearly 100\% \cite{atlas2016photon}. Pseudo-rapidity regions are selected between -2.5 and 2.5 for jet and between -1.44 and 1.44 for photon based on Fig. \ref{fig:eta_dist}.

%\subsection{\label{subsec:eta} Pseudo-Rapidity Distributions  }
%\subsection{\label{subsec:pT} Transverse Momentum Distributions  }

 Invariant mass cut is applied as $\text{m}^{\star}-2\Gamma^{\star}<\text{m}_{\gamma j}<\text{m}^{\star}+2\Gamma^{\star}$ mass window, here  m$^{\star}$	indicates excited quark mass and $\Gamma^{\star}$ denotes total decay width of corresponding excited quark.  Additionally, to differentiate photon and jet, cone angle radius ($\Delta R$) cut is used as $\Delta R > 0.5$. 
  
  \begin{equation}
 \label{eq:significance}
 S_{stat}=\frac{\sigma_{s}}{\sqrt{\sigma_{s}+\sigma_{b}}} \sqrt{\mathcal{L}_{i n t}}
 \end{equation}
 
 To obtain statistical significance values, Eq. \ref{eq:significance} is utilized, where $\sigma_{s}$ depicts signal cross section, $\sigma_{b}$ means background cross section and $\mathcal{L}_{i n t}$ is integrated luminosity. By using this formula, all confidence levels, discovery ($5\sigma$), observation ($3\sigma$) and exclusion ($2\sigma$) are computed with applied cuts as mentioned above  for FCC and S$pp$C colliders.  
 
\section{Results and Conclusions \label{sec:conc}}

\begin{figure*}
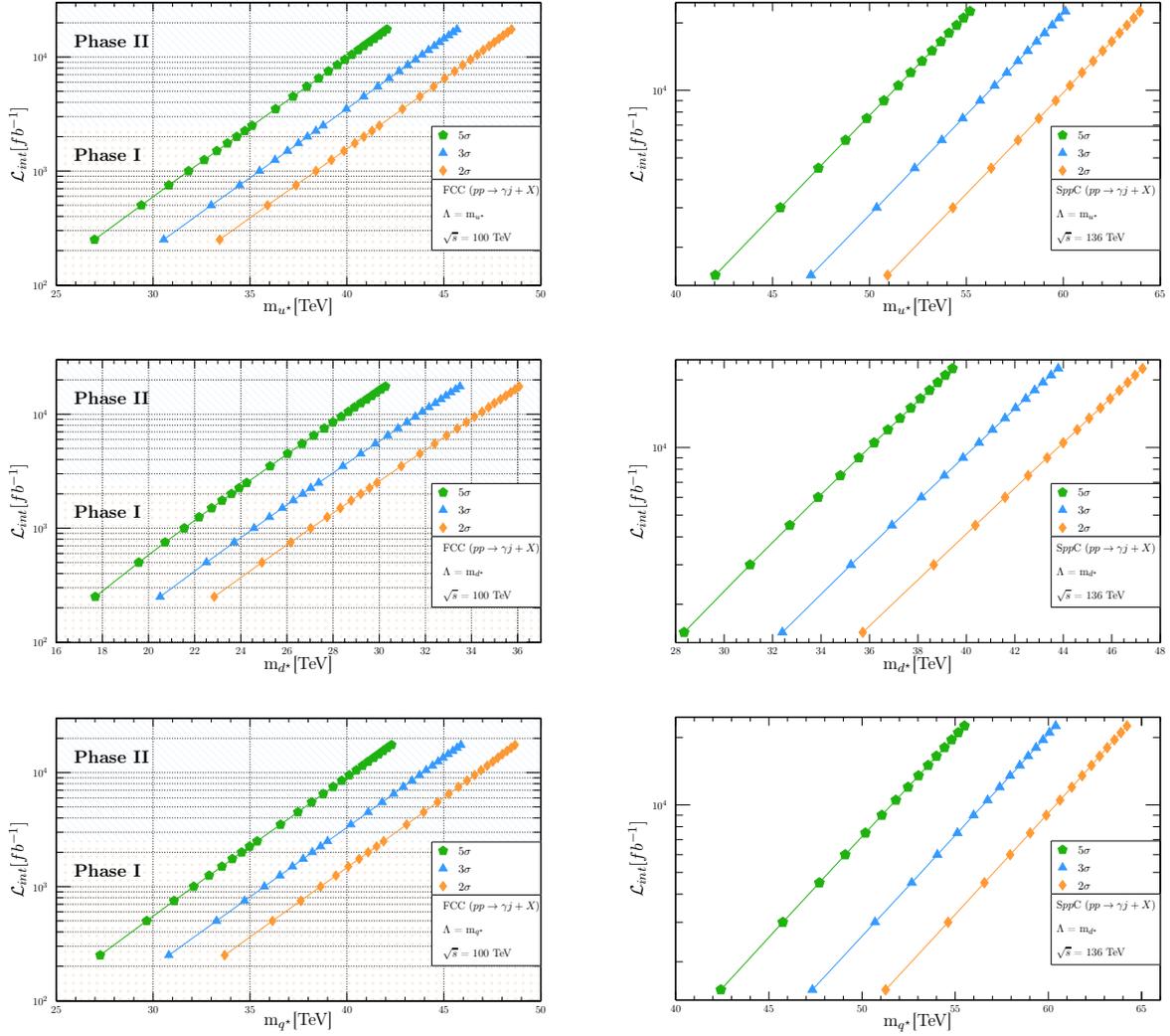

	\scalebox{0.405}{\input{LumiMass_FCC_ustar_LEM_Aj.tex}}	
	\scalebox{0.405}{\input{LumiMass_SppC_ustar_LEM_Aj.tex}}	
	\scalebox{0.405}{\input{LumiMass_FCC_dstar_LEM_Aj.tex}}	
	\scalebox{0.405}{\input{LumiMass_SppC_dstar_LEM_Aj.tex}}	
	\scalebox{0.405}{\input{LumiMass_FCC_qstar_LEM_Aj.tex}}	
	\scalebox{0.405}{\input{LumiMass_SppC_qstar_LEM_Aj.tex}}		
	\caption{Luminosity mass  plots for the FCC (left panel) and S$pp$C (right panel).  \label{fig:lumiMass}}
\end{figure*}

In this analysis, excited quark mass limits depending on  integrated luminosity is calculated by implementing cut values  and utilizing statistical significance formula mentioned in previous section for both collider options. These limits are presented in Fig. \ref{fig:lumiMass} based on expected operation time of FCC and S$pp$C colliders.  Excited quarks' discovery ($5\sigma$), observation ($3\sigma$) and exclusion ($2\sigma$) mass limits are listed in Tab. \ref{tab:lumiMass}, which reveals both colliders' physics potentials in terms of final luminosity values. At the end of  FCC-Phase I operational time, excited quarks discovery limit will be at 35.4 TeV. FCC-Phase II will make this discovery limit further up to 42.3 TeV. On the other hand, S$pp$C with 22500 fb$^{-1}$ luminosity will take the discovery limits of Q$^\star$ up to 55.5 TeV.  

\begin{table*}[b]
	\caption{Attainable mass limits of excited quarks at FCC and S$pp$C corresponding to  integrated luminosity values by selecting $\Lambda$ = m$_{Q^{\star}}$. \label{tab:lumiMass}}
%	\resizebox{\textwidth}{!}{
\begin{ruledtabular}
			\begin{tabular}{lcccccccccc}
			\multicolumn{2}{l}{\textbf{Colliders :}}&  \multicolumn{3}{c}{\textbf{FCC-Phase I}} &  \multicolumn{3}{c}{\textbf{FCC-Phase II}} & \multicolumn{3}{c}{\textbf{S$pp$C}} \\ 
			\hline 
			\multicolumn{2}{l}{\textbf{$\mathcal{L}_{\textbf{int}}$ } (fb$^{-1}$) :}&  \multicolumn{3}{c}{\textbf{2500}} &  \multicolumn{3}{c}{\textbf{17500}} & \multicolumn{3}{c}{\textbf{22500}} \\ 
			\hline 
			\multicolumn{2}{l}{\textbf{Significance :}} &\textbf{$5\sigma$} &\textbf{$3\sigma$ }& \textbf{$2\sigma$}&\textbf{$5\sigma$} &\textbf{$3\sigma$ }& \textbf{$2\sigma$}&\textbf{$5\sigma$} &\textbf{$3\sigma$ }& \textbf{$2\sigma$} \\ 
		%	\cline{1-2}
			%\multicolumn{2}{|p{3cm}|}{\textbf{Excited Quark Mass} (TeV)}&  & &  &  &  &  &  &   &  \\
			\hline 
			\textbf{m$_{u^{\star}}$}& & 35.1  & 38.8 & 41.7 & 42.1 & 45.7 & 48.5 &55.2  & 60.1  & 64.0 \\ 
			\cline{1-1} \cline{3-11}
				\textbf{m$_{d^{\star}}$} & (TeV) & 24.3& 27.4 & 29.9 & 30.3  & 33.5 & 36.1 & 39.4  & 43.8  & 47.3 \\ 
			\cline{1-1}  \cline{3-11}
			\textbf{m$_{q^{\star}}$} &   & 35.4  & 39.0 & 41.9   & 42.3  & 45.9  & 48.7  & 55.5  & 60.4 & 64.2 \\ 
		\end{tabular} 
		\end{ruledtabular} 
%	}
\end{table*}

\begin{figure*}
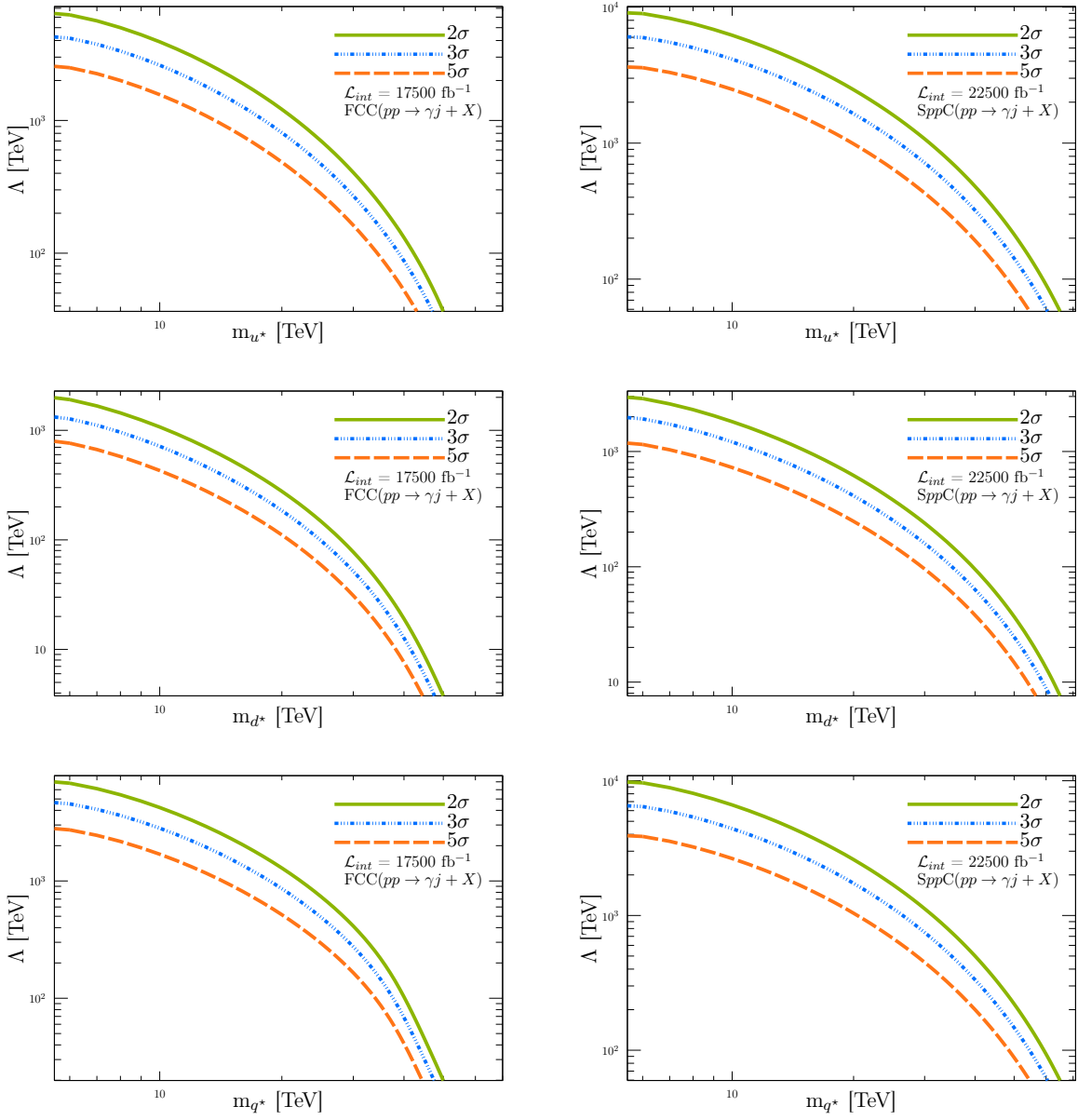

	\scalebox{0.4}{\input{LamdaMass_FCC_Aj_ustar_Lumi17500.tex}}	
	\scalebox{0.4}{\input{LamdaMass_SppC_Aj_ustar_Lumi22500.tex}}	
	\scalebox{0.4}{\input{LamdaMass_FCC_Aj_dstar_Lumi17500.tex}}	
	\scalebox{0.4}{\input{LamdaMass_SppC_Aj_dstar_Lumi22500.tex}}	
	\scalebox{0.4}{\input{LamdaMass_FCC_Aj_qstar_Lumi17500.tex}}	
	\scalebox{0.4}{\input{LamdaMass_SppC_Aj_qstar_Lumi22500.tex}}		
	\caption{Compositeness scale-mass  plots for the FCC (left panel) and S$pp$C (right panel).  \label{fig:lamdaMass}}
\end{figure*}

Determining compositeness scale value is an important matter that could be decouple of the excited quark mass, so this parameter can be scanned starting from experimental mass limit of excited quark. Figure \ref{fig:lamdaMass} provides attainable compositeness scale values of excited quarks at the both energy frontier machines with their final luminosities. As can be seen from Fig. \ref{fig:lamdaMass},  degenerate cases held the highest compositeness scale values if the m$_{q^\star}$ = 5.5 TeV: 2798 TeV (5$\sigma$),  4663 TeV  (3$\sigma$)   and  6995 TeV (2$\sigma$) at the FCC with $\mathcal{L}_{int}$ = 17500 fb$^{-1}$ and 3913 TeV (5$\sigma$),  6521 TeV  (3$\sigma$)   and  9781 TeV (2$\sigma$) at the S$pp$C with $\mathcal{L}_{int}$ = 22500 fb$^{-1}$. For some selected mass quantities, attainable compositeness scale values are listed in Tab. \ref{tab:LamdaSearchFCC} and \ref{tab:LamdaSearchSppC}. 

\begin{table}[h!]
	\caption{Achievable compositeness scale values for selected mass quantities for all three cases at FCC with final integrated luminosity values. \label{tab:LamdaSearchFCC}}
	\begin{ruledtabular}
	\begin{tabular}{ccccccccccc}

		\multicolumn{2}{c}{\multirow{3}{1cm}{\textbf{Mass (TeV)}}} & \multicolumn{9}{c}{\textbf{Compositeness Scale $\Lambda$} (TeV) \textbf{at FCC}} \\ \cline{3-11}
		\multicolumn{2}{c}{ }& \multicolumn{3}{c|}{\textbf{$u^{\star}$}} &   \multicolumn{3}{c|}{\textbf{$d^{\star}$}}   &   \multicolumn{3}{c}{\textbf{$q^{\star}$}} \\ \cline{3-11}
		\multicolumn{2}{c}{ }& \textbf{$5\sigma$ }&\textbf{$3\sigma$ } &\textbf{$2\sigma$ } &\textbf{$5\sigma$ } &\textbf{$3\sigma$ } &\textbf{$2\sigma$}  & \textbf{$5\sigma$} & \textbf{$3\sigma$} & \textbf{$2\sigma$} \\ \hline
		\multicolumn{2}{c}{\textbf{5.5}}& 2559       & 4265       & 6398  &796      & 1327       & 1991 & 2798       & 4663      & 6995\\	\hline
		\multicolumn{2}{c}{\textbf{10}} &  1564       & 2607      & 3911   & 429       & 716       & 1074  & 1690       & 2817      & 4226 \\ \hline
		\multicolumn{2}{c}{\textbf{20}} &485       & 808       & 1213 & 111       & 185       & 278  & 516       & 859       & 1289  \\ \hline
		\multicolumn{2}{c}{\textbf{30}}& 162       & 270      & 406  & 31       & 52      & 77  &165       & 275       & 412\\

	\end{tabular} 
\end{ruledtabular}
	%	}
\end{table}

\begin{figure*}[h!]
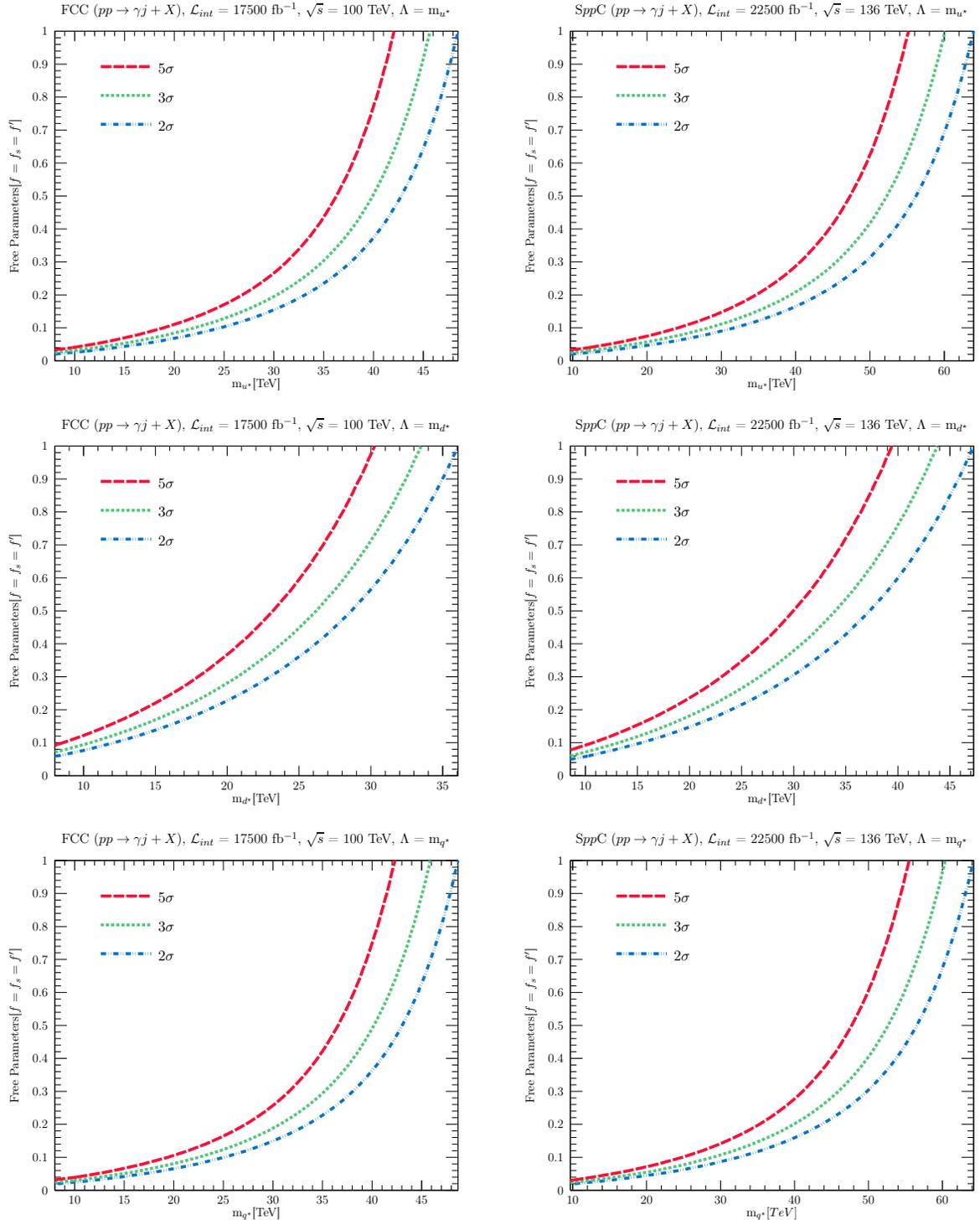

	\scalebox{0.4}{\input{ustar_FCC_LEM_Aj_f_par_search.tex}}	
	\scalebox{0.4}{\input{ustar_SppC_LEM_Aj_f_par_search.tex}}	
	\scalebox{0.4}{\input{dstar_FCC_LEM_Aj_f_par_search.tex}}	
	\scalebox{0.4}{\input{dstar_SppC_LEM_Aj_f_par_search.tex}}	
	\scalebox{0.4}{\input{qstar_FCC_LEM_Aj_f_par_search.tex}}	
	\scalebox{0.4}{\input{qstar_SppC_LEM_Aj_f_par_search.tex}}		
	\caption{Free parameters scan for excited quarks at FCC (left panel) anf at S$pp$C (right panel).  \label{fig:fSearch}}
\end{figure*}

\begin{table}[h!]
	\caption{Achievable compositeness scale values for selected mass quantities for all three cases at S$pp$C with final integrated luminosity values. \label{tab:LamdaSearchSppC}}
	\begin{ruledtabular}
		\begin{tabular}{ccccccccccc}

			\multicolumn{2}{c}{\multirow{3}{1cm}{\textbf{Mass (TeV)}}} & \multicolumn{9}{c}{\textbf{Compositeness Scale $\Lambda$} (TeV) \textbf{at S$pp$C}} \\ \cline{3-11}
			\multicolumn{2}{c}{ }& \multicolumn{3}{c|}{\textbf{$u^{\star}$}} &   \multicolumn{3}{c|}{\textbf{$d^{\star}$}}   &   \multicolumn{3}{c}{\textbf{$q^{\star}$}} \\ \cline{3-11}
			\multicolumn{2}{c}{ }& \textbf{$5\sigma$ }&\textbf{$3\sigma$ } &\textbf{$2\sigma$ } &\textbf{$5\sigma$ } &\textbf{$3\sigma$ } &\textbf{$2\sigma$}  & \textbf{$5\sigma$} & \textbf{$3\sigma$} & \textbf{$2\sigma$} \\ \hline
			\multicolumn{2}{c}{\textbf{5.5}}&  3631       & 6051       & 9077 &  1183       & 1972       & 2958  &  3913       & 6521       & 9781\\	\hline
			\multicolumn{2}{c}{\textbf{10}} &   2485       & 4142        & 6213  &   726       & 1210       & 1815  &   2646       & 4411       & 6616 \\ \hline
			\multicolumn{2}{c}{\textbf{20}}&   988       & 1646       & 2469 &   248        & 413       & 620  &    1040       & 1733       & 2599   \\	\hline
			\multicolumn{2}{c}{\textbf{30}} & 430       & 716        & 1074& 96       & 160       & 239 &   449       & 748       & 1122 \\ \hline
			\multicolumn{2}{c}{\textbf{40}}&   192       & 321       & 481  &   38       & 63       & 95  &   200       & 334       & 500 \\
			
		\end{tabular} 
	\end{ruledtabular}
	%	}
\end{table}
\newpage
Up to now, free parameters were taken as $f=f_{s}=f^{\prime}=1$; however, these parameters could have different values which can be lower than 1. Mass limits of excited quark at all confidence level are presented  in Fig. \ref{fig:fSearch}, using both colliders' final luminosity values and taking $\Lambda$ = m$_{Q^\star}$ for different coupling free parameters from around 0.1 up to 1.  It is clearly seen in Fig. \ref{fig:fSearch},  even free parameters are taken 0.1, degenerate case excited quark, $q^\star$,  could be discovered at 19.4 TeV at FCC and 24.8 TeV at S$pp$C. 

\newpage
The presented results show that FCC and S$pp$C will have significant physics potential  as new frontier machines. They will potentially discover excited quarks up to 42.3 TeV at FCC and 55.5 TeV at S$pp$C. If  excited quarks are discovered  at these colliders,  compositeness scale of excited quarks will be determined.  

\newpage

\section*{ACKNOWLEDGMENTS}
	This study is supported by TUBITAK under the grant No {[}114F337{]}.
	Authors are grateful to the Usak University, Energy, Environment and
	Sustainability Application and Research Center for their support.
	We also wish to acknowledge Assist. Prof. Yunus Emre Akbana for language
	editing and Dr. Hasan Eskalen for technical support.

\section*{\label{sec:ref}References}

\bibliographystyle{apsrev4-1}
\bibliography{excited_quark}

\end{document}